\documentclass[11pt,a4paper]{article}
\usepackage[utf8]{inputenc}
\usepackage{amsmath}
\usepackage{mathtools}
\usepackage{a4wide}
\usepackage{cite}
\usepackage{authblk}
\usepackage{graphicx}
\usepackage{amsmath,amssymb}
\usepackage[enable]{easy-todo}
\usepackage[hang,small,bf]{caption}
\usepackage[subrefformat=parens]{subcaption}
\usepackage{easy-todo}
\usepackage{url}

\title{Low-scale leptogenesis and dark matter \\
in a three-loop radiative seesaw model}
\author{Osamu Seto$^{a}$, Tetsuo Shindou$^{b}$, and Takanao Tsuyuki$^{b}$\footnote{tsuyuki@cc.kogakuin.ac.jp}}
\affil{\small $^a$\textit{Department of Physics, Hokkaido University, Sapporo 060-0810, Japan}}
\affil{\small $^b$\textit{Division of Liberal-Arts, Kogakuin University, Hachioji, Tokyo 192-0015, Japan}} 
\date{}
\begin{document}

\maketitle
\begin{abstract}
We show that three open questions in particle physics and cosmology: the origin of neutrino mass, the identity of dark matter, and the origin of the baryon asymmetry of the universe can be explained simultaneously in the three-loop seesaw model proposed by Krauss, Nasri, and Trodden.
We discuss the difficulty of successful leptogenesis with three right-handed neutrinos, 
and we propose a scenario with four right-handed neutrinos that 
satisfies all observational constraints. 
This scenario predicts a sleptonlike particle as light as a few 
hundreds GeV that can be probed by future collider experiments.
\end{abstract} 

\begin{flushleft}
EPHOU-22-022, KU-PH-033
\end{flushleft}

\section{INTRODUCTION}

The observation of neutrino oscillation has confirmed that neutrinos have masses. 
The neutrino oscillation parameters~\cite{Esteban:2020cvm} 
and the cosmological observations~\cite{Planck:2018vyg}
indicate that neutrino masses are many orders of magnitude smaller than 
those of the other standard model (SM) fermions. 
We need some mechanism beyond the SM to explain such tiny masses. 

When we consider cosmology, there are additional strong motivations to consider the new physics beyond the SM. 
One is that there is no appropriate candidate for dark matter (DM) in the SM. 
Another problem is that the baryogenesis does not work in the SM, as the electroweak baryogenesis requires a smaller Higgs boson mass than the observed one. 

The canonical seesaw mechanism~\cite{Minkowski:1977sc,Yanagida:1979as,Gell-Mann:1979vob,Mohapatra:1979ia} 
is a favorable idea to address the origin of the neutrino masses, 
in which neutrino masses are suppressed by heavy right-handed (RH) neutrino masses.
In such models, on the one hand, the thermal leptogenesis~\cite{Fukugita:1986hr} works as a mechanism of the baryogenesis, 
where the $CP$-violating decay of the right-handed neutrino produces the lepton asymmetry, which is partially converted to the baryon asymmetry through the sphaleron process. 
The disadvantage of the thermal leptogenesis in the seesaw model is that it tends to require a right-handed neutrino to be as heavy as $10^9$~GeV~\cite{Davidson:2002qv,Buchmuller:2002rq,Giudice:2003jh} and cannot be tested by experiments. 
In addition, further extensions would be necessary because the minimal seesaw model does not contain a suitable candidate for DM. 

An alternative approach to explain the tininess of the neutrino masses is to utilize loop factors~\cite{Zee:1980ai,Cheng:1980qt,Zee:1985id,Babu:1988ki,Krauss:2002px,Ma:2006km,Aoki:2008av}. 
A class of models with right-handed neutrinos, so-called radiative seesaw models, is particularly attractive. 
In these models, a discrete symmetry under which the right-handed neutrinos and some extra scalars 
are odd is introduced to forbid the tree-level neutrino-mass generation. 
This symmetry can also stabilize the lightest odd-charged particle as the DM.

In this paper, we focus on the radiative seesaw model proposed by Krauss, Nasri, and Trodden~\cite{Krauss:2002px}  (often called the KNT model), 
where tiny neutrino masses are generated via three-loop diagrams. 
To forbid the tree-level contribution to neutrino masses, 
a $Z_2$ symmetry is introduced under which the right-handed neutrinos and a charged 
scalar $S_2$ is odd. 
Therefore, the lightest right-handed neutrino can be a candidate for the DM. 
The phenomenology of the KNT model has been studied in Refs.~\cite{Cheung:2004xm,Ahriche:2013zwa,Chowdhury:2018nhd,Cepedello:2020lul,Irie:2021obo,Seto:2022ebh}.
In our previous work~\cite{Seto:2022ebh}, we have found that the KNT model is severely constrained for the inverted neutrino-mass ordering case by the experiments searching for the lepton flavor violation (LFV).
A model proposed by Ma~\cite{Ma:2006km} (sometimes called the scotogenic model) also shares properties that neutrino masses are radiatively generated at the one-loop level and possess a dark matter candidate. 
From the viewpoint of the philosophy of radiative generation of neutrino mass, the KNT model would be more appealing than Ma's scotogenic model with a TeV scale mass of extra scalars that need a small scalar quartic coupling $\mathcal{O}(10^{-5})$ besides the one-loop suppression factor that is not enough to reduce neutrino masses down to sub-eV scale. Phenomenologically, the DM in the KNT model is a Majorana right-handed neutrino, while that in the scotogenic model is the scalar in a heavier inert doublet Higgs doublet.\footnote{The case of the lightest right-handed neutrino dark matter is hardly compatible with various experimental results and theoretical consistency~\cite{Lindner:2016kqk} or needs fine-tuned mass spectrum so that co-annihilation works~\cite{Suematsu:2009ww}.} There are different dark matter phenomenology. 

Baryogenesis in the context of the KNT model has been scarcely considered. 
The high-scale leptogenesis in an extended KNT model has been studied~\cite{Gu:2016xno} but not for the original KNT model. 
In the KNT model, the decay of the second lightest right-handed neutrino $N_2$ into a charged lepton $\ell_R^\mp$ and $S_2^\pm$ can produce the lepton asymmetry. 
One difficulty of the leptogenesis in the KNT model is that the asymmetry generated by the $N_2$ decay is cancelled and washed out\footnote{This problem does not exist in the scotogenic model with scalar DM} after the decay $S_2^{\pm} \rightarrow N_1 \ell_R^{\pm}$. 
In this paper, we point out that this washout can be suppressed if $S_2$ is relatively light and its Boltzmann suppression is not strong at the sphaleron freeze-out time. 
Since the sphaleron freeze-out temperature $T_{\mathrm{sph}}$ is about $\simeq 130$~GeV~\cite{DOnofrio:2014rug}, $S_2$ must be lighter than a few hundreds GeV. 
The light $S_2$ behaves like a slepton and can be explored by direct searches at the collider experiments such as the International Linear Collider (ILC)~\cite{Behnke:2013xla,NunezPardodeVera:2022izz}, the Compact Linear Collider~\cite{Aicheler:2018arh}, the Future Circular Collider~\cite{FCC:2018evy}, or Circular Electron Positron Collider (CEPC)~\cite{CEPCStudyGroup:2018ghi,Yuan:2022ykg}.
Another difficulty of the leptogenesis is due to the $\Delta L=2$ washout processes
 such as $\ell_i^\pm S_2^\mp \leftrightarrow \ell_j^\mp S_2^\pm$ via exchange of right-handed neutrinos~\cite{Ma:2006fn,Haba:2011ra}. 
Those washout reaction rates are, in general, many order of magnitude larger than the cosmic expansion rate, hence, the generated lepton asymmetry hardly survive. 
Thus we may need an extension of the model to explain the baryon asymmetry via 
leptogenesis.
A simple solution is introducing the fourth generation of the right-handed neutrino, 
and we adopt this possibility in this paper. 
We check that our scenario of leptogenesis is compatible with the observations such as the neutrino oscillation, the DM abundance, and LFV processes.

This paper is organized as follows. In Sec.~\ref{sec:KNTmodel}, we briefly review the KNT model and discuss the constraints from the neutrino oscillation, the lepton flavor violations, 
the relic abundance of the dark matter, and direct searches of new particles. 
In Sec.~\ref{sslep}, we discuss the possibility of leptogenesis, 
and we show that the observed baryon asymmetry can be explained in the case of four 
right-handed neutrinos.
In Sec.~\ref{sec:summary}, we summarize this paper and make concluding remarks. 

\section{The KNT model}
\label{sec:KNTmodel}
\subsection{The Lagrangian and the neutrino-mass matrix}

We consider the KNT model~\cite{Krauss:2002px}, which explains the tininess of neutrino masses by utilizing the loop factor. 
In the model, charged scalar fields $S_1$ and $S_2$ and RH neutrinos $N_I$ with $I$ are the generation indices are introduced. 
Furthermore, a global $Z_2$ symmetry is introduced, under which $S_1$ and $N_I$ fields are assigned odd and $S_2$ is assigned even.

This symmetry is necessary to forbid the neutrino Yukawa couplings with the right-handed neutrinos, left-handed lepton doublets, and the Higgs doublet, which would provide too large Dirac neutrino-mass terms after the electroweak symmetry breaking. 
The $Z_2$ symmetry simultaneously guarantees the stability of the lightest $Z_2$ odd particle. If $N_1$ is the lightest, it can be a DM candidate. 
The Lagrangian terms added to the SM are
\begin{align}
  \mathcal{L}_{\text{KNT}}=
\frac{h_{ij}}{2}\overline{L_i^c}i\tau_2 L_j S_1^+ 
+ g_{Ij}^*\overline{N_I^c}\ell_{Rj}S_2^+
+\frac{m_{N_I}}{2}\overline{N_{I}^c}N_{I}+\mathrm{H.c.}-V,
\label{lagKNT}
\end{align}
where the superscript $c$ denotes the charge conjugation, the Yukawa matrix $(h_{ij})$ is an antisymmetric, {\it i.e.}, $h_{ij}=-h_{ji}$, $g_{Ij}$ are other Yukawa coupling constants, $\lambda_S$ is a complex coupling, and $N_I$ are in the mass basis.
The scalar potential $V$ includes four-point scalar coupling terms,
\begin{align}
V&\supset \frac{\lambda_S}{4}(S_1^-)^2(S_2^+)^2+\mathrm{H.c.}.
\end{align}

\begin{figure}
\begin{center}
\includegraphics{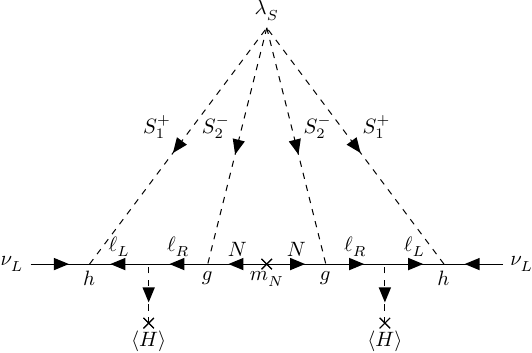}
\end{center}
\caption{The diagram of the neutrino-mass generation in the KNT model.}
\label{Fig:MnuInKNTmodel}
\end{figure}

\begin{figure}
\begin{center}
\includegraphics[width=12cm]{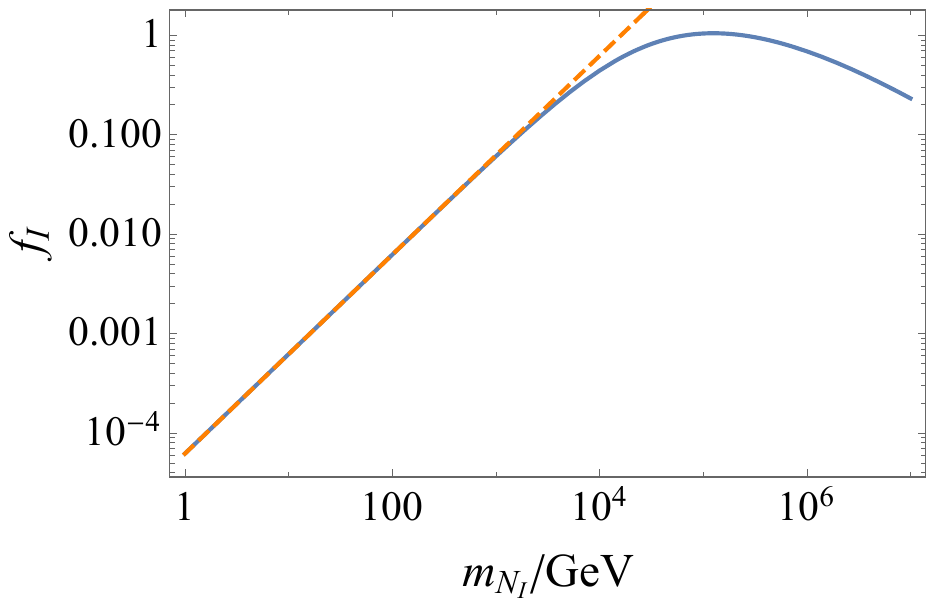}
\end{center}
\caption{
    The numerical behavior of the loop function $f_I$.
    The blue line shows the loop function $f_I$ in the expression of the neutrino masses in Eq.~(\ref{eq:KNTneutrinomassmatrix}) 
    with $m_{S_1}=2.3\times 10^{4}$~GeV, $m_{S_2}=100$~GeV. 
    The orange-dashed line shows the analytic approximation given in Eq.~(\ref{ef1}).}
\label{floop}
\end{figure}

With the Lagrangian, the neutrino masses are induced through the Feynman diagram in Fig.~\ref{Fig:MnuInKNTmodel}. 
A component of the neutrino-mass matrix is given by~\cite{Ahriche:2013zwa}\footnote{Our convention is slightly different from Ref.~\cite{Ahriche:2013zwa}: $h_{ij}=f_{ij}/2$ and $f_I=\sqrt{y}F(x_I,y)$.}
\begin{align}
M_{ab} &=
\frac{\lambda_{S}}{4(4\pi)^3m_{S_1}}\sum_{I,j,k}m_{\ell_j}m_{\ell_k}h_{aj}h_{bk}g_{Ij}
g_{Ik}f_I\;. 
\label{eq:KNTneutrinomassmatrix}
\end{align}
Here, we use a simple expression for the loop function $f_I$ found in Ref. \cite{Seto:2022ebh}:
\begin{equation}
    \begin{aligned}
    f_I&=\frac{\sqrt{x_I}}{8y^{3/2}}\int_0^\infty dr \frac{J^2}{r(r+x_I)},\\
    J&=q\ln\left[\frac{y}{q}\right]+\frac{y}{q}\ln[q]+(1+r)\ln\left[\frac{1+r}{y}\right],\\
	q &= \frac{1}{2}\left(1+r+y+\sqrt{(1+r+y)^2-4y}\right),\\
    x_I &= \frac{m_{N_I}^2}{m_{S_2}^2},\ y = \frac{m_{S_1}^2}{m_{S_2}^2}.
    \end{aligned}
    \label{eq:fi}
\end{equation}
As discussed later, since observations constrain 
$m_{N_1}\lesssim \mathcal{O}(100)\text{ GeV}$, $m_{S_1}\gtrsim \mathcal{O}(10^{4})\text{ GeV}$, and $m_{S_2}\simeq \mathcal{O}(10^{2})\text{ GeV}$, the region of $y\gg x_1$ and $y\gg 1$ is of interest. 
In such a case, the loop function can be analytically estimated as 
\begin{align}
f_1\simeq \frac{\zeta(2)+\zeta(3)}{2}\sqrt{\frac{x_1}{y}}\simeq 1.42\frac{m_{N_1}}{m_{S_1}},
\label{ef1}
\end{align}
where $\zeta$ is the Riemann zeta function. 
In Fig.~\ref{floop}, we plot the loop function in the case of $y\gg 1$ and the analytic expression equation~(\ref{ef1}), and we find that the analytic expression provides a good approximation in the region $m_{N_1}\lesssim 0.1 m_{S_2}$.

To find a parameter set that reproduces the neutrino oscillation data by the mass matrix equation~(\ref{eq:KNTneutrinomassmatrix}), we can use the relations found in Refs.~\cite{Irie:2021obo,Seto:2022ebh}. 
In Eq. (\ref{eq:KNTneutrinomassmatrix}), there are terms suppressed by the small electron mass $m_e$, which can be ignored. 
Under this approximation, we can extract conditions on $h_{ij}$ as
\begin{align}
\frac{M_{e\mu}M_{\mu\tau}-M_{e\tau}M_{\mu\mu}}{M_{\mu\mu}M_{\tau\tau}-M_{\mu\tau}^2} &=  \frac{h_{12}}{h_{23}},\label{ek}\\
\frac{M_{e\mu}M_{\tau\tau}-M_{e\tau}M_{\mu\tau}}{M_{\mu\mu}M_{\tau\tau}-M_{\mu\tau}^2}&= \frac{h_{13}}{h_{23}}.\label{ekp}
\end{align}

\subsection{Lepton flavor violation}

In the KNT model, the charged lepton flavor is not conserved. Thus, charged leptons 
$\ell_i$ can decay into the lighter one $\ell_j$ and a photon $\gamma$. 
The branching ratio of this process is estimated as~\cite{Ahriche:2013zwa,Seto:2022ebh}
\begin{align}
\mathrm{Br}(\ell_i\to \ell_j\gamma)
&=\frac{48\pi^3\alpha_{\mathrm{em}}}{G_F^2}\left(|A_L^{ij}|^2+|A_R^{ij}|^2\right)\mathrm{Br}(\ell_i\to \ell_j\nu\bar{\nu}), 
\label{ebr0}
\end{align}
where $A_R^{ij}$ and $A_L^{ij}$ are given by 
\begin{equation}
    \begin{aligned}
        A_R^{ij}=&\frac{1}{16\pi^2m_{S_2}^2}\sum_{I=1}^{n_N}g_{Ii}^*g_{Ij}F_2(x_I), \label{ear}\\
        A_L^{ij}=&
        \frac{1}{192\pi^2m_{S_1}^2}h_{il}h_{jl}^*\quad (l\neq i,j),
    \end{aligned}
\end{equation}
$\alpha_\mathrm{em}$ is the fine structure constant, and $G_F$ is the Fermi constant.
The loop function $F_2(x)$ is defined 
as~\cite{Bertolini:1990if}\footnote{
  This function $F_2(x)$ differs from the $F_1(x)$ in Ref.~\cite{Chowdhury:2018nhd} by factor 2, i.e., $F_2(x)=\frac{1}{2}F_1(x)$.
} 
\begin{equation}
  F_2(x)= \frac{2x^2+5x-1}{12(x-1)^3}-\frac{x^2\log(x)}{2(x-1)^4}\;.
\end{equation}

The most severe constraint comes from $\mu\to e \gamma$. 
The current upper bound on this process is given by 
Br($\mu\to e \gamma$)$<4.2\times 10^{-13}$~\cite{MEG:2016leq}.
Even if $A_R^{21}=0$ is satisfied by taking a specific form for $g_{Ii}$, 
$A_L^{21}$ cannot be taken as zero because of the antisymmetric structure of $(h_{ij})$. 
Since $|A_R^{21}|^2\geq 0$, the branching ratio satisfies 
\begin{equation}
    \begin{aligned}
        \mathrm{Br}(\mu\to e\gamma)
        &\geq\frac{48\pi^3\alpha_{\mathrm{em}}}{G_F^2}\left|\frac{h_{23}h_{13}^*}{192\pi^2m_{S_1}^2}\right|^2\\
        &=2.22\times 10^{-14}|h_{23}|^4\left|\frac{h_{13}}{h_{23}}\right|^2 \left(\frac{10^4\mathrm{GeV}}{m_{S_1}}\right)^4.
    \end{aligned}
    \label{ebrm}
\end{equation}
Note that the factor $h_{13}/h_{23}$ is determined by the elements of the neutrino-mass matrix, as shown in Eq.~(\ref{ekp}). 

As for neutrino-mass parameters, we input the best-fit values of normal ordering with super-Kamiokande data in Ref.~\cite{Esteban:2020cvm}. 
In the KNT model, the lightest active neutrino mass is zero because of the antisymmetric structure of $(h_{ij})$~\cite{Irie:2021obo}. Hence, the other active neutrino masses are determined by the observed mass-squared differences. 
In this case, the remaining free parameter in $M_{ab}$ is one Majorana phase. 
In the following, we set the Majorana phase to zero for simplicity.
With a finite value of the Majorana phase, the analysis does not change much. 
If the neutrino-mass ordering is inverted, the factor $|h_{13}/{h_{23}|}$ is larger than that of the normal ordering case~\cite{Irie:2021obo}. 
The combination of the upper bound on Br($\mu\to e\gamma$) and the perturbativity condition was studied in Ref.~\cite{Seto:2022ebh}, 
and inverted ordering is severely constrained.

By inputting the neutrino oscillation data and the upper bound of 
Br($\mu\to e \gamma$) to Eq.(\ref{ebrm}), we obtain the lower bound on $m_{S_1}$:
\begin{align}
m_{S_1}> 8700 |h_{23}| \ \mathrm{GeV}.  
\label{ems1lfv}
\end{align}

\subsection{Slepton searches}
In the KNT model, $S_2$ behaves like a purely right-handed slepton in supersymmetric models. 
In the case considered below ($g_{11}=g_{12}=0$), it is like a right-handed stau $\tilde{\tau}_R$. 
The mass bounds on $\tilde{\tau}_R$ are obtained by the LEP experiments~\cite{
ALEPH:2001oot,ALEPH:2003acj,DELPHI:2003uqw,L3:2003fyi,OPAL:2003nhx}.\footnote{In the analysis by the CMS Collaboration \cite{CMS:2022rqk}, 
constraints on the degenerate or purely left-handed stau were obtained. 
However, they note that their sensitivity for the purely right-handed stau was insufficient and the constraint was not available.}
The bounds come from the search for the stau decay ($S_2\to \tau+N_1$ in our case) 
and they depend on $m_{N_1}$. In the exclusion plot in Ref.~\cite{LEP}, the strongest bound is $m_{S_2}\gtrsim 95.5$~GeV (95\% confidence level) at $m_{N_1}\simeq 64$~GeV. 
In the analysis below, we use a conservative bound
\begin{align}
m_{S_2} > 96 \text{ GeV} \label{ems2lo}
\end{align}
for all the $m_{N_1}$ region. 

\subsection{Dark matter}
\label{ssdm}
The lightest right-handed neutrino $N_1$ is stabilized by the $Z_2$ symmetry and it can be the dark matter. 
We assume that $N_1$ was produced as thermal relics. The abundance of such dark matter is determined by the annihilation cross section. 
Since the dark matter mass $m_{N_1}$ cannot be very large, 
we naively expect to have a significant contribution to the LFV via 
$S_2$ and $N_1$ exchange diagrams. 
To avoid such a contribution, it is preferred that 
$N_1$ couples to only one lepton flavor.
Thus, we here make an ansatz that $N_1$ only couples to $\tau$, i.e., 
$g_{11}=g_{12}=0$ and $g_{13}\neq 0$.
With this ansatz, 
the cross section is calculated as~\cite{Krauss:2002px,Cheung:2004xm,Ahriche:2013zwa}
\begin{align}
\langle\sigma v\rangle \simeq \frac{m_{N_1}^2(m_{N_1}^4+m_{S_2}^4)}{8\pi (m_{N_1}^2+m_{S_2}^2)^4}
  |g_{13}|^4 \frac{1}{x_f}
  \;, \label{edm}
\end{align}
with $x_f\simeq 20$~\cite{Kolb:1990vq}. 
The relic abundance of the dark matter after the decoupling is approximated by
\begin{align}
\Omega_{N_1}h^2\simeq 0.12\frac{2.9\times 10^{-9}\text{ GeV}^{-2}}{\langle \sigma v \rangle}. 
\end{align}
By comparing the observed dark matter abundance $\Omega_{N_1} h^2=0.120\pm 0.001$ \cite{Planck:2018vyg}, we obtain
\begin{align}
|g_{13}| &\simeq 0.35\frac{x_1+1}{[x_1(x_1^2+1)]^{1/4}} \left(\frac{m_{S_2}}{100 \text{ GeV}}\right)^{1/2}, \label{eg13}
\end{align}
where $x_1=m_{N_1}^2/m_{S_2}^2$.

\begin{figure}
  \begin{center}
  \includegraphics[width=14cm]{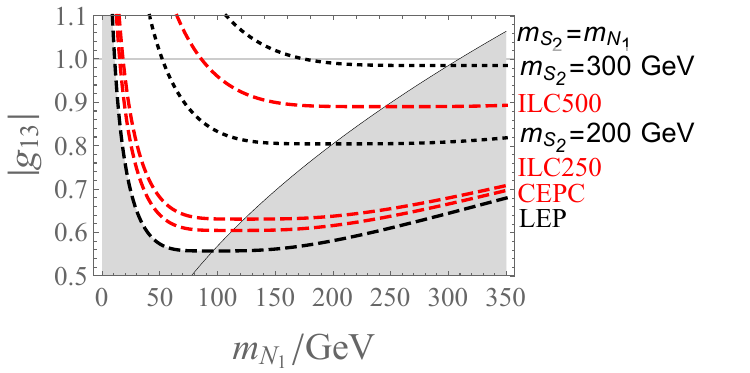}
  \end{center}
  \caption{Contour of $m_{S_2}$ in the plane of the dark matter mass $m_{N_1}$ 
  and its Yukawa coupling $|g_{13}|$. 
  The white region satisfies conditions to explain dark matter and leptogenesis.
  The black-dashed curve shows the simplified lower bound from LEP ($m_{S_2}>96$~GeV)~\cite{LEP} 
  and red-dashed curves show the simplified sensitivity up limit on $m_{S_2}$ of future experiments (CEPC: 113~GeV~\cite{Yuan:2022ykg}, ILC250: 123~GeV~, ILC500: 245~GeV~\cite{NunezPardodeVera:2022izz}). 
  The dotted lines are contours of $m_{S_2}=200\ \text{GeV},\ 300\ \text{GeV}$.
  These curves are obtained by using the DM condition equation (\ref{eg13}). 
  The right-bottom shaded corner is the region where the charged $S_2$ is lighter than $N_1$. 
  The horizontal-gray line shows the perturbativity up limit.}
  \label{fig:g13DM}
\end{figure}

In Fig.~\ref{fig:g13DM}, we show the contour of $m_{S_2}$ in the $m_{N_1}$-$|g_{13}|$ plane which can reproduce the thermal relic abundance of the DM.
By the perturbativity condition $|g_{13}|\leq 1$, there is a lower and upper bound on the dark matter mass $m_{N_1}$:
\begin{align}
11 \text{ GeV} < m_{N_1} < 310 \text{ GeV}. \label{emn1}
\end{align}
The upper bound gives
\begin{align}
m_{S_2} < 310 \text{ GeV}.
\end{align}
As discussed later, $m_{S_2}$ less than $\mathcal{O}(100)~\text{GeV}$ is also preferred to explain the baryon asymmetry by the leptogenesis.

\subsection{The minimal case}
We here consider the minimal structure of $(g_{Ii})$ to 
explain both the neutrino mixing and the dark matter with 
satisfying the lepton flavor violation constraints. 

To explain the neutrino oscillation data, we should reproduce 
$M_{\tau\tau}$, $M_{\mu\mu}$, and $M_{\mu\tau}$.
Once these elements are reproduced, the other elements can be tuned by $h_{12}/h_{23}$ and $h_{13}/h_{23}$.
To realize it, we need at least three independent $g_{Ij}$.
On the other hand, the elements $g_{I1}$ are irrelevant to the neutrino-mass matrix because of the strong suppression by $m_e$. 
Therefore, we need at least two RH neutrinos which have a significant size of 
the Yukawa coupling with $\mu$ and/or $\tau$.
If we use $N_1$ and $N_3$ for generating an appropriate neutrinomass matrix,\footnote{
    As discussed in Sec.~\ref{sslep}, 
    $N_2$ will be necessary for the leptogenesis, 
    and the mass $m_{N_2}$ is required to be 
    smaller than $m_{N_3}$. So that we here use $N_3$ instead of $N_2$.
} 
$g_{32}$ and $g_{33}$ should have significant size, 
as $g_{22}=0$ is taken in Sec.~\ref{ssdm}.
In addition, $g_{31}$ should be small to avoid too large a contribution to 
$\mu\to e \gamma$ through the $S_2$ exchange diagrams.
Thus, the minimal setup of the Yukawa couplings
with $\overline{N_I^c}\ell_{Ri}S_2^+$ 
for the neutrino mixing and the dark matter is given by 
\begin{align}
    \mathcal{L}=
    \begin{pmatrix} 
        \overline{N_1^c}&
        \overline{N_3^c}
    \end{pmatrix}
    \begin{pmatrix}
        0&0&g_{13}^*\\
        0&g_{32}^*&g_{33}^*
    \end{pmatrix}
    \begin{pmatrix}
        e_R\\
        \mu_R\\
        \tau_R
    \end{pmatrix}
    S_2^+
    +\text{h.c.}\;.
\end{align}


With the setup, three components of the neutrino-mass matrix in Eq.~(\ref{eq:KNTneutrinomassmatrix}) become
\begin{align}
M_{\mu\mu} &= \frac{\lambda_S m_\tau^2 h_{23}^2 }{4(4\pi)^3 m_{S_1}}(g_{13}^{2}f_1+g_{33}^2f_3)\;, \label{emmumu}\\
M_{\mu\tau} &= -\frac{\lambda_S m_\mu m_\tau h_{23}^2 }{4(4\pi)^3 m_{S_1}} g_{32}g_{33} f_3\;, \label{emmut}\\
M_{\tau\tau} &= \frac{\lambda_S m_\mu^2 h_{23}^2 }{4(4\pi)^3 m_{S_1}}g_{32}^{2} f_3\;,
\label{emtt}
\end{align}
and we can see that there are enough degrees of freedom to reproduce an appropriate 
neutrino matrix. 
Note that the inverted ordering case with the best-fit oscillation parameters is excluded by the
$\mu\to e\gamma$ constraint.\footnote{
It corresponds to the $n_{\text{eff}}=1$ case in Ref.~\cite{Seto:2022ebh}.}
Thus, we consider the normal ordering case throughout this paper. 

Let us comment on the lepton flavor violation processes other than $\mu\to e\gamma$.
If the constraint (\ref{ems1lfv}) is satisfied, the contribution of $S_1$ to the other decays 
$\tau\to e \gamma$ and $\tau \to \mu \gamma$ is the same order as $\mu\to e\gamma$
and much weaker than the experimental constraints. 
In our case of $g_{I1}=0$, $S_2$ does not contribute to $\mu\to e \gamma$ and $\tau\to e \gamma$.
On the other hand, 
the $S_2$ contribution to $\tau\to \mu \gamma$ is given as
\begin{align}
\text{Br}(\tau\to \mu\gamma)&\simeq\frac{48\pi^3\alpha_{em}}{G_F^2}\left|\frac{g_{32}^*g_{33}}{16\pi^2 m_{S_2}^2}F_2(x_3)\right|^2\text{Br}(\tau\to \mu\nu\bar{\nu}). \label{ebrt}
\end{align}
Using the upper bound Br($\tau\to \mu \gamma$)$<4.2\times 10^{-8}$ \cite{Belle:2021ysv} and Br$(\tau\to \mu\nu\bar{\nu})=0.1739$ \cite{ParticleDataGroup:2020ssz}, we find
\begin{align}
|g_{32}g_{33}|F_2(x_3)<2.7\times 10^{-3}\left(\frac{m_{S_2}}{100\text{ GeV}}\right)^2. \label{etmg}
\end{align}

\section{Leptogenesis} \label{sslep}
\subsection{Production and evolutions of the asymmetry}
\begin{figure}
    \begin{center}
    \includegraphics[scale=0.9]{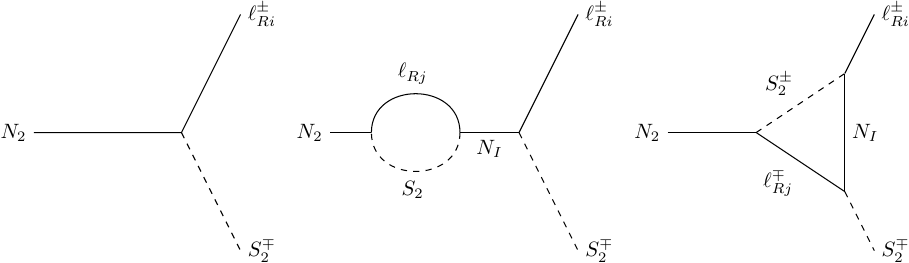}
    \end{center}
    \caption{
        The Feynman diagrams relevant to the $CP$ asymmetric decay 
        of $N_2\to \ell_RS_2$ at the one-loop level.
    }
    \label{CPasymN2decay}
\end{figure}

For baryogenesis, we consider a scenario that the $CP$-violating decay 
of $N_2$ to $\ell_R$ and $S_2$ generates the lepton asymmetry, and 
the lepton asymmetry is converted to the baryon asymmetry by the sphaleron process.
The $CP$ asymmetry in the decay $N_2\to \ell_{Ri}^{\pm}S_2^{\mp}$ is defined as 
\begin{equation}
    \epsilon_i \equiv \frac{\Gamma(N_2\to \ell_{Ri}^{-}S_2^+)-\Gamma(N_2\to \ell_{Ri}^+S_2^-)}{\Gamma(N_2\to \ell_{Ri}^{-}S_2^+)+\Gamma(N_2\to \ell_{Ri}^+S_2^-)}\;,
\end{equation}
which comes from the interference between the tree diagram and the loop diagrams shown in 
Fig.~\ref{CPasymN2decay}.

For producing enough large lepton asymmetry, 
the decay width of $N_2$ should not be too large compared to the Hubble rate
at around $T=m_{N_2}$.
To examine this, 
we define $K$ as the ratio of the total decay width of $N_2$, $\Gamma_{N_2} = \Gamma(N_2\to \ell_R^++S_2^-)+\Gamma(N_2\to \ell_R^-+S_2^+)$ 
and the Hubble rate $H$ at $T=m_{N_2}$ \cite{Kolb:1990vq}:
\begin{equation}
        K \equiv\left.\frac{\Gamma_{N_2}}{2H}\right|_{T=m_{N_2}} 
        =\frac{\sum_i|g_{2i}|^2}{8\pi}m_{N_2}\times\left(\frac{8\pi^3g_*}{90}\right)^{-\frac{1}{2}}\frac{M_P}{m_{N_2}^2}
        =2.8\times 10^{13} \sum_i|g_{2i}|^2 \frac{10^3\text{ GeV}}{m_{N_2}}\;,
    \label{ekdef}
\end{equation}
where $g_*=110.5$ is the effective relativistic degrees of freedom by taking into account $S_2$, $N_1$, and $N_2$, and 
$K\lesssim 1$ is required.
To realize it, the Yukawa couplings $g_{2i}$ should be too strongly suppressed 
for $N_2$ to contribute to the neutrino-mass matrix. 
Thus $N_2$ should be introduced in addition to $N_1$ and $N_3$.

We consider the chemical potentials before the sphaleron freeze-out to derive the relation between the baryon number and the lepton number in our model. 
We denote the chemical potentials by $\mu$ and subscripts $q,u,d,L,\ell_R, \phi$  indicate 
the SM quark doublet, right-handed up-type quark, right-handed down-type quark, 
lepton doublet, right-handed charged lepton, and the Higgs doublet. 
The chemical potential of the Majorana particles is zero $\mu_N=0$. 
The SM Yukawa interactions and the sphaleron process yields\footnote{We ignore the effects of the top quark decoupling and phase transition near the sphaleron freeze-out temperature.}
\begin{align}
-\mu_q-\mu_\phi+\mu_u &=0, \\
-\mu_q+\mu_\phi+\mu_d &=0, \\
-\mu_{L}+\mu_\phi+\mu_{\ell_R} &=0, \\
3\mu_q+\mu_L &= 0.
\end{align}
The conditions from vanishing hypercharge in the Universe and the non-SM interaction are~\cite{Gu:2016xno}
\begin{align}
3(\mu_q-\mu_d+2\mu_u-\mu_L-\mu_{\ell_R})+2(\mu_\phi+\mu_{S_2})&=0,\\
\mu_{S_2}+\mu_{\ell_R}&=0.
\end{align}
These equations can be solved as
\begin{equation}
    \begin{aligned}
        \mu_q &= -\frac{8}{9}\mu_{\ell_R},\
        \mu_u =\frac{7}{9}\mu_{\ell_R},\ 
        \mu_d = -\frac{23}{9}\mu_{\ell_R},\\ 
        \mu_L &=\frac{8}{3}\mu_{\ell_R},\ 
        \mu_\phi = \frac{5}{3}\mu_{\ell_R},\ \mu_{S_2}=-\mu_{\ell_R}.
    \end{aligned}
\end{equation}
Now, the condition (\ref{ems2up}) becomes clear. If the condition is not satisfied, it ends up with $\mu_{S_2}=-\mu_{\ell_R}=0$.
The relation of the baryon number $B=3\times 3\times (1/3)\times (2\mu_q+\mu_u+\mu_d)$ and the lepton number $L=\sum_i(\mu_{L_i}+\mu_{\ell_{Ri}})$ is expressed as
\begin{align}
B=\frac{32}{79}(B-L). 
\label{ebbl}
\end{align}

The sets of Boltzmann equations to be solved are those of $N_2, \ell_{i}$ and $S_2^\pm$.
Those are rewritten as the Boltzmann equations for the number density of $N_2$ and asymmetries of number densities of $ B/3-L_i$ and $S_2^+-S_2^-$, respectively, as ($n_X$ denotes the number density of the species $X$)
\begin{align}
 \frac{d n_{N_2}}{dt} + 3 H n_{N_2} =& - \sum_i\langle \Gamma(N_2 \rightarrow S_2^\pm \ell_i^\mp)\rangle ( n_{N_2} - n^{\mathrm{eq}}_{N_2}) \nonumber \\ 
 &-\sum_{i, j}\langle \sigma v(N_2 N_1 \leftrightarrow \ell_i^\mp \ell_j^\pm)\rangle (n_{N_2}n_{N_1}-n^{\mathrm{eq}}_{N_2}n^{\mathrm{eq}}_{N_1}) \nonumber \\
 &-\sum_{i, j}\langle \sigma v(N_2 \ell_i^\mp \leftrightarrow \ell_j^\mp N_1)\rangle(n_{N_2}n_{\ell_i^\mp}-n^{\mathrm{eq}}_{N_2}n^{\mathrm{eq}}_{\ell_1^\mp})  , \label{Eq:BoltN_n} \\
 \frac{d n_{B/3-L_i}}{dt} + 3 H n_{B/3-L_i} =&- \epsilon_i \langle \Gamma(N_2 \rightarrow S_2^\pm \ell_i^\mp)\rangle (n_{N_2}-n^{\mathrm{eq}}_{N_2})  \nonumber \\
& - \langle \Gamma(S_2^+ \ell_i^- \rightarrow N_2 )\rangle n_{B/3-L_i} - \langle \Gamma(S_2^+ \ell_i^- \rightarrow N_2 )\rangle n_{S_2^+ - S_2^-} \nonumber \\
& +\langle \Gamma(S_2^+ \rightarrow N_1 \ell_i^+)\rangle n_{S_2^+ - S_2^-}  - \frac{1}{2} \langle \Gamma(N_1 \ell_i^+ \rightarrow S_2^+)\rangle n_{B/3-L_i} \nonumber \\
 & -\sum_j \langle\sigma v (\ell_i^- \ell_j^-\leftrightarrow S_2^- S_2^-) \rangle \left(\frac{n_{\ell_i^- +\ell_i^+}}{2} n_{B/3-L_j} +\frac{n_{\ell_j^-+\ell_j^+} }{2} n_{B/3-L_i} \right) \nonumber \\
 &+\sum_j \langle\sigma v (\ell_i^+ \ell_j^+\leftrightarrow S_2^+ S_2^+) \rangle n_{S_2^+ + S_2^-} n_{S_2^+ - S_2^-} \nonumber \\
 &+\sum_j \langle\sigma v (\ell_i^- S_2^+\leftrightarrow \ell_j^+ S_2^-) \rangle \left(\frac{n_{\ell_i^- +\ell_i^+}}{2} n_{S_2^+ - S_2^-} -\frac{n_{S_2^+ + S_2^-}}{2} n_{B/3-L_i} \right)  \nonumber \\
 &+\sum_j \langle\sigma v (\ell_i^+ S_2^-\leftrightarrow \ell_j^- S_2^+) \rangle\left(\frac{n_{\ell_j^- +\ell_j^+}}{2} n_{S_2^+ - S_2^-} -\frac{n_{S_2^+ + S_2^-}}{2} n_{B/3-L_j}\right) , \\
 \frac{d n_{S_2^+-S_2^-}}{dt} + 3 H n_{S_2^+-S_2^-} =& \sum_i\epsilon_i\langle\Gamma(N_2 \rightarrow S_2^\pm \ell_i^\mp)\rangle (n_{N_2}-n^{\mathrm{eq}}_{N_2}) \nonumber \\
 & +\sum_i\langle \Gamma(S_2^+ \ell_i^- \rightarrow N_2 )\rangle n_{S_2^+-S_2^-} + \sum_i\langle \Gamma(S_2^+ \ell_i^- \rightarrow N_2 )\rangle n_{B/3-L_i}\nonumber \\
 & - \sum_i\langle \Gamma(S_2^+ \rightarrow N_1 \ell_i^+)\rangle n_{S_2^+ - S_2^-}+ \sum_i\frac{1}{2}\langle \Gamma(N_1 \ell_i^+ \rightarrow  S_2^+ )\rangle  n_{B/3-L_i} \nonumber \\
 &+\sum_{i, j} \langle\sigma v (\ell_i^- \ell_j^-\leftrightarrow S_2^- S_2^-) \rangle \left(\frac{n_{\ell_i^- +\ell_i^+ }}{2} n_{B/3-L_j}  +\frac{n_{\ell_j^- +\ell_j^+ } }{2} n_{B/3-L_i}  \right) \nonumber \\
 &-\sum_{i, j} \langle\sigma v (\ell_i^+ \ell_j^+\leftrightarrow S_2^+ S_2^+) \rangle n_{S_2^+ + S_2^-} n_{S_2^+ - S_2^-} \nonumber \\
 &-\sum_{i, j} \langle\sigma v (\ell_i^- S_2^+\leftrightarrow \ell_j^+ S_2^-) \rangle \left(\frac{n_{\ell_i^- +\ell_i^+}}{2} n_{S_2^+ - S_2^-} -\frac{n_{S_2^+ + S_2^-}}{2} n_{B/3-L_i}   \right)  \nonumber \\
 &-\sum_{i, j} \langle\sigma v (\ell_i^+ S_2^-\leftrightarrow \ell_j^- S_2^+) \rangle\left(\frac{n_{\ell_j^- +\ell_j^+}}{2} n_{S_2^+ - S_2^-} -\frac{n_{S_2^+ + S_2^-}}{2} n_{B/3-L_j}  \right),
\end{align}
 where $\langle\Gamma\rangle$ and $\langle\sigma v\rangle$ are thermal averaged decay (inverse decay) rates and thermal averaged scattering cross section times relative velocity, respectively.

It is a kind of thermal leptogenesis but has some remarkable features. 
First, 
the sum of $B-L$ asymmetry and the asymmetry between $S_2^+$ and $S_2^-$ is always zero,
i.e. $\sum_in_{B/3-L_i}+n_{S_2^+-S_2^-}=0$,
which is nothing but the electric charge neutrality of the Universe. 
If $S_2$ becomes nonrelativistic and decouples from the thermal equilibrium before the
sphaleron freeze-out, the abundance of $S_2$ gets Boltzmann suppressed and the lepton 
asymmetry is washed out. 
However, after the sphaleron decouples from the thermal bath at 
the temperature $T_{\mathrm{sph}}=131.7\pm 2.3$~GeV~\cite{DOnofrio:2014rug}, 
the baryon asymmetry $n_B/s$ is not washed out anymore, 
while $B-L$ asymmetry is decreasing. 
With Eq.~(\ref{ebbl}), the final baryon asymmetry is given by
\begin{align}
    \frac{n_B}{s}=\frac{32}{79}\left.\frac{n_{B-L}}{s}\right|_{T=T_{\mathrm{sph}}}.
    \end{align}
We stress that the out-of-equilibrium condition for successful baryogenesis in this scenario is satisfied by the sphaleron decoupling.
It is different from the situation in the canonical leptogenesis. 
As a consequence of the behavior, 
the mass of $S_2$ should not be much larger than $T_{\mathrm{sph}}$,
so that 
\begin{align}
m_{S_2} < \mathcal{O}(100) ~\mathrm{ GeV}. 
\label{ems2up}
\end{align}

Second, we discuss the strong washout effect of the scattering process. 
$\Delta L=2$ washout processes $\ell_i^\pm S_2^\mp \leftrightarrow \ell_j^\mp S_2^\pm$
and $\ell_i^\pm \ell_j^\pm \leftrightarrow S_2^\pm S_2^\pm$ via $N_I$ exchange are, in 
general, very strong~\cite{Ma:2006fn}. 
Those cross sections are of the order of $\langle \sigma v\rangle \sim |g_{Ii}^*g_{Ij}|^2/T^2$.
The interaction rate $\Gamma \sim T^3 \langle \sigma v\rangle$ is much larger than 
the Hubble parameter $H \sim T^2/M_P$ where $M_P=1.22\times 10^{19}$~GeV 
is the Planck mass. 
This means that the $\Delta L=2$ washout is very strong unless involved 
$g$ are extremely small. 
Since the strength of $\Delta L=2$ washout are flavor dependent,
we need to evaluate lepton asymmetry in each flavor. 
To reproduce the observed neutrino mixing in the KNT model, 
$g_{32}$ and $g_{33}$ cannot be small, so that 
the $\mu$ and $\tau$ asymmetries are strongly washed out.
Producing $e$ asymmetry by the $CP$-violating decay of  
$N_2\to e_{R}^{\pm}S_2^{\mp}$ is only the possibility. 
The $CP$-violating parameter of $N_2\to e S_2$ decay is 
evaluated as 
\begin{equation}
    \begin{aligned}
        \epsilon_1&=\frac{1}{4\pi} \sum_{i=3,4}\frac{\text{Im}[(gg^\dag)_{2i}^2]}{(gg^\dag)_{11}}F(m_{N_i}^2/m_{N_2}^2)  ,\\
        F(x)&\equiv x^{1/2}\left(1+(1+x)\ln\frac{x}{1+x}+\frac{1}{1-x}\right).
    \end{aligned}
\end{equation}
In order to produce $e$ asymmetry in the model with three RH neutrinos, a
significant size of $g_{31}$ is required to provide the large $CP$ violation 
in $N_2\to e_R^{\pm}S_2^{\mp}$ decay. 
However, such a large $g_{31}$ is disfavored 
not only by the $\mu\to e\gamma$ constraint
but also by the strong washout effect via 
the flavor changing $\Delta L=2$ scattering 
$e^{\pm}S_2^{\mp} \leftrightarrow \ell_i^{\mp}S_2^{\pm}$ $(\ell_i=\mu, \tau)$. 
In fact, we cannot find any point which reproduces 
enough large baryon asymmetry in the case with three RH neutrinos.

A simple extension to solve the above difficulty is 
introducing the fourth RH neutrino $N_4$.
The neutrino-mass matrix and the properties of the DM 
can be explained by $N_1$ and $N_3$, while 
$N_2$ and $N_4$ play an important role in the leptogenesis.
In the case that $N_2$ and $N_4$ only couple to an electron, 
$\epsilon_1$ can be enhanced by large $g_{41}$, 
while
the contribution to $\mu\to e\gamma$ via $N_4$ and $N_2$ exchange and 
the washout by the flavor changing $\Delta L=2$ scattering processes
are absent. 

\subsection{Benchmark inputs}
We construct a benchmark scenario to demonstrate how the 
baryon asymmetry is produced in our leptogenesis scenario. 
With the four RH neutrinos, the Yukawa matrix $g$ is a $4\times 3$ matrix.
The minimal structure for the successful leptogenesis is given by 
\begin{equation}
    g=\begin{pmatrix}
        0&0&g_{13}\\
        g_{21}&0&0\\
        0&g_{32}&g_{33}\\
        g_{41}&0&0
    \end{pmatrix}\;.
\end{equation}
With this structure, there are nine complex and six real parameters:
\begin{equation}
    \begin{aligned}
        \text{complex}:&\ \lambda_S,\ h_{12},\ h_{13},\ h_{23},\ g_{13},\ g_{21},\ g_{32},\ g_{33},\ g_{41}\\
        \text{real}:&\ m_{S_1},\ m_{S_2},\ m_{N_1},\ m_{N_2},\ m_{N_3},\ m_{N_4}.
    \end{aligned}
\end{equation}
These parameters have to satisfy 
the neutrino-mass conditions, Eqs. (\ref{ek}), (\ref{ekp}), (\ref{emmumu}), (\ref{emmut}), 
and (\ref{emtt}),
and two more conditions, i.e.,  leptogenesis
and the DM conditions.
The four inequalities (\ref{ems1lfv}), (\ref{etmg}), (\ref{ems2lo}), and (\ref{ems2up}), 
and the perturbativity condition (couplings have to be smaller than the order of unity) 
must also be obeyed. 


In our analysis, 
we scan $m_{N_2}$ and $m_{S_2}$ in the range of 
$[100,330]$~GeV, 
and we fix the other parameters to satisfy the 
conditions obtained from the DM relic abundance and neutrino oscillation data.
First, we consider the DM relic abundance.
Once $|g_{13}|$ and $m_{S_2}$ are fixed, 
$m_{N_1}$ is determined to reproduce the relic abundance of the DM
as shown in Fig.~\ref{fig:g13DM}.
For example, $m_{S_2}=110$~GeV and $g_{13}=1.0$ give $m_{N_1}=12$~GeV. 

Second, we take into account the neutrino oscillation data.
For the neutrino-mass parameters,
we input the best-fit neutrino oscillation parameters in the case with super-Kamiokande data \cite{Esteban:2020cvm}.
Once we fix $h_{23}=1$, 
Eqs.~(\ref{ek}) and (\ref{ekp}) determine 
the rest of the $h_{ij}$ as 
\begin{align}
h_{12} &= 0.600e^{-0.0480i}, \ h_{13} = 0.329 e^{0.102i}.
\end{align}
We additionally fix $\lambda_S=|g_{32}|=1$. 
By using the neutrino-mass equations.~(\ref{emmumu}), (\ref{emmut}), and (\ref{emtt}) and 
the approximation formula~(\ref{ef1}) of $f_1$, we obtain
\begin{align}
M_{\mu\mu}-\frac{M_{\mu\tau}^2}{M_{\tau\tau}} &=\frac{1.42 m_\tau^2 m_{N_1}}{4(4\pi)^3 m_{S_1}^2}g_{13}^{2}, \\
M_{\tau\tau} &= \frac{m_\mu^2}{4(4\pi)^3m_{S_1}}g_{32}^{2}f_3, \\
\frac{M_{\mu\tau}}{M_{\tau\tau}} &= -\frac{m_\tau}{m_\mu}\frac{g_{33}}{g_{32}}.
\end{align}
These equations have a solution 
\begin{equation}
    \begin{aligned}
        g_{13} &= 1.0,\     g_{32}=1.0,\ g_{33} = -0.053,\\
        m_{S_1}&= 2.33\times 10^4 \text{ GeV},\ m_{N_3} = 3.67\times 10^6 \text{ GeV}.
    \end{aligned}
\end{equation}
The branching ratios of the LFV decays in the benchmark point are evaluated as 
\begin{align}
\text{Br}(\mu\to e\gamma)
&=8.2\times 10^{-17},\\
\text{Br}(\tau\to \mu\gamma)
&=7.5\times 10^{-15},
\end{align}
which are far below the current limits.

Finally, we fix the rest of the parameters relevant to the leptogenesis.
For optimizing the production of the lepton asymmetry, 
we tune the value of $|g_{21}|$ to satisfy $K=1$ in Eq.~(\ref{ekdef}) \textit{i.e.},
\begin{equation}
    |g_{21}|=1.9\times 10^{-7}\left(\frac{m_{N_2}}{10^3~\mathrm{GeV}}\right)^{1/2}\;, 
\end{equation}
and we take $\arg(g_{21})=\pi/4$ which maximize the $CP$ asymmetry $\epsilon_1$.
We fix the $m_{N_4}$ and $g_{41}$ as $m_{N_4}=1.0\times 10^8$~GeV and $g_{41}=0.1$, respectively.
Our benchmark inputs are summarized in Table~\ref{tabbench}. 
\begin{table}
    \caption{Definition of benchmark inputs.}\label{tabbench}
    \begin{center}
    \begin{tabular}{c|c}
        Parameter& Value\\ \hline\hline
        $m_{S_1}$ & $2.33\times 10^4$~GeV\\ \hline
        $m_{S_2}$ & Scanned in $[100, 330]$~GeV\\ \hline
        $m_{N_1}$ & Depending on $m_{S_2}$\\ \hline
        $m_{N_2}$ & Scanned in $[100, 330]$~GeV\\ \hline
        $m_{N_3}$ & $3.67\times 10^{6}$~GeV\\ \hline
        $m_{N_4}$ & $1.0\times 10^8$~GeV\\ \hline
        $\lambda_S$& $1.0$\\\hline 
        $(h_{12}, h_{23}, h_{13})$& 
        $(0.600e^{-0.0480i},\  1.0,\  0.329 e^{0.102i})$\\ \hline
        $(g_{13}, g_{32}, g_{33}, g_{41})$& $(1.0,\ 1.0,\ -0.053, \ 0.1)$\\ \hline
         $|g_{21}|$ & Depending on $m_{N_2}$\\ \hline
        $\arg(g_{21})$ & $\pi/4$\\ \hline
    \end{tabular}
\end{center}
\end{table}

\subsection{Numerical analysis}
\begin{figure}
  \begin{center}
  \includegraphics[width=12cm]{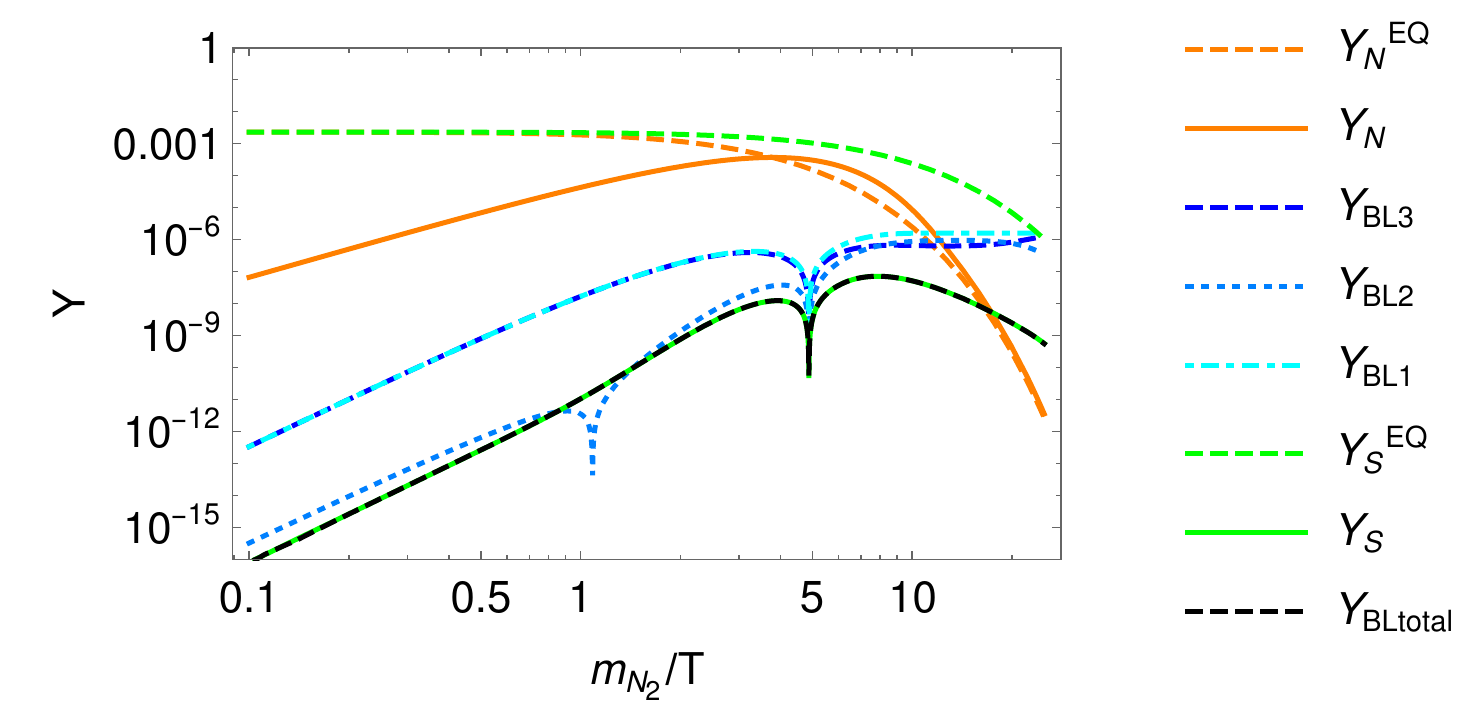}
  \end{center}
  \caption{The evolution of yields of each species indicated by the subscriptions: $N,\ \text{BL}i,\ S,\ \text{BLtotal}$ label yields of $N_2$, $B/3-L_i$, $S_2$, $B-\sum_i L_i$, respectively. The superscript ``EQ'' indicates that the line is in thermal equilibrium.
  The yields are calculated by the Boltzmann equations with the parameters $m_{S_2}=110\text{ GeV}, m_{N_2}=250\text{ GeV}$, and the other parameters are shown in Table~\ref{tabbench}.}
  \label{fig:Yevolution}
\end{figure}

We show, in Fig.~\ref{fig:Yevolution}, 
the evolution of absolute values of asymmetry yield $Y \equiv |n_i|/s$ of each leptons
with bluish (dotted) dashed curves, that of $S_2^+ - S_2^-$ with the green curve, 
that of $N_2$ with the orange curve, 
in the case with $m_{S_2}=110$~GeV and $m_{N_2}=250$~GeV.
The asymmetry of the $N_2$ decay rates $\epsilon_1$ is \cite{Covi:1996wh}:
which reflects the charge neutrality, as mentioned above.
The total $B-L$ asymmetry drawn with the black dashed curve always coincides with that of $S_2^+ - S_2^-$ asymmetry.
We can see that both the total $B-L$ asymmetry and the $S_2^\pm$ asymmetry 
decreases for a large $m_{N_2}/T$.
At $T=T_{\mathrm{sph}}$, the baryon asymmetry is frozen out
as $Y_B=\left.Y_{B-L}\right|_{T=T_{\mathrm{sph}}}$ 
due to the sphaleron decoupling.
Figure~\ref{fig:Yevolution} shows that enough large 
baryon asymmetry $Y_B=\mathcal{O}(10^{-10})$ is obtained
in our benchmark. 

\begin{figure}
    \begin{center}
    \includegraphics[width=10cm]{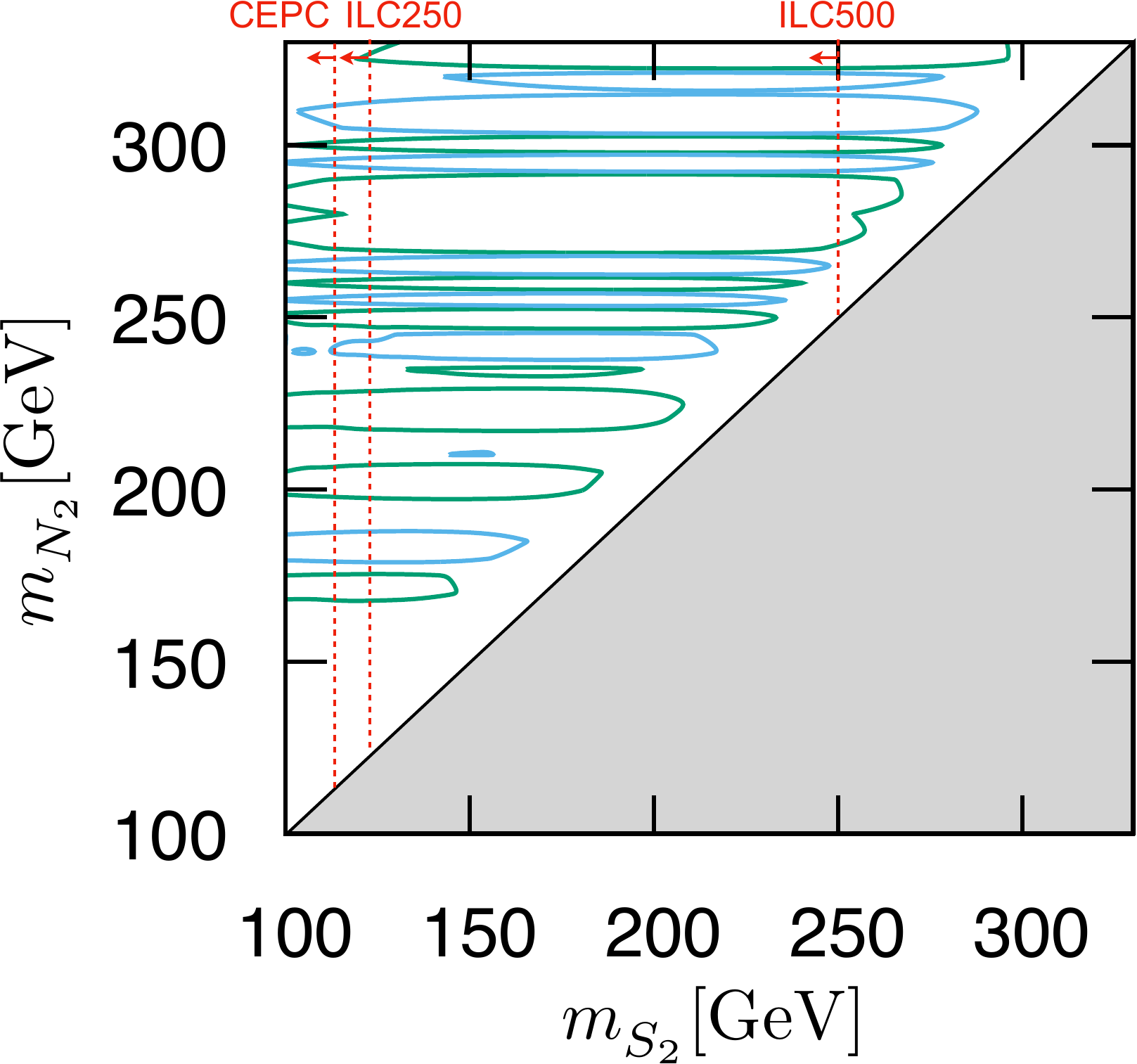}
    \end{center}
    \caption{Contours of the yield $Y_B:=n_B/s$ for a given set of Yukawa coupling and the $CP$ violation in the $m_{S_2}$-$m_{N_2}$ plain. Inside of the blue lines is the region of $Y_B>10^{-10}$ and inside of the green lines gives $Y_B<-10^{-10}$. 
    The vertical-dashed lines show the sensitivities of future experiments: CEPC, ILC at $\sqrt{s}=250\text{ GeV}$ and $\sqrt{s}=500\text{ GeV}$. 
    }
    \label{fig:mSmN}
  \end{figure}

In Fig.~\ref{fig:mSmN}, 
we show an example of contour plots of the final baryon asymmetry for a set of 
Yukawa coupling constants.
The range $m_{S_2}<310\text{ GeV}$ is preferred by the DM relic abundance as shown in Fig.~\ref{fig:g13DM} and 
it can be explored by future $e^+e^-$ collider 
experiments.
For example, the CEPC can probe it up to $113$~GeV~\cite{Yuan:2022ykg},
and the ILC with the center of mass energy of $250$~GeV can do up to 
$123$~GeV~\cite{NunezPardodeVera:2022izz}. The Compact Linear Collider with $380$~GeV or 
the ILC with $500$~GeV can explore our predicted mass range.


%

\section{Summary}
\label{sec:summary}

In this paper, we have proved that neutrino masses, dark matter, and leptogenesis can be explained in the KNT model with four right-handed neutrinos. 
We have shown explicit parameters that can realize these phenomena and satisfy the observational constraints, such as the LFV decays. 
To avoid the severe constraint from the LFV decay on the inverted mass ordering case \cite{Seto:2022ebh}, we considered the normal ordering case. 
In our scenario, there is a definite prediction that a charged scalar particle $S_2$ should be as light as $m_{S_2}<\mathcal{O}(100)$~GeV. 
The $S_2$ in our scenario behaves as a staulike particle, 
and it can be detected by future lepton collider experiments. 

\section*{Acknowledgments}
The authors are grateful to Takashi Toma for valuable comments. This work is supported in part by the Japan Society for the Promotion of Science 
(JSPS) KAKENHI Grants No. 20H00160 (T.S.), Grants No. JP19K03860, No. JP19K03865, No. 23K03402 and MEXT KAKENHI Grant No. 21H00060 (O.S.).

\bibliography{ref}
\bibliographystyle{utphys}
\end{document}